%% file: main.tex
\documentclass[manuscript]{acmart}
\usepackage{xspace}
\usepackage{graphicx}
\usepackage{bm}
\usepackage{marvosym}
\usepackage{subcaption}
\usepackage{color}
\usepackage{colortbl}
\usepackage{multirow}

\definecolor{myblack}{RGB}{0,0,0}
\newcommand{\paratitle}[1]{\vspace{1.5ex}\noindent\textbf{#1}}
\newcommand{\ie}{\emph{i.e.,}\xspace}

\newcommand{\eg}{\emph{e.g.,}\xspace}

\newcommand{\ignore}[1]{}

\AtBeginDocument{%
  \providecommand\BibTeX{{%
    \normalfont B\kern-0.5em{\scshape i\kern-0.25em b}\kern-0.8em\TeX}}}

\setcopyright{acmcopyright}
\copyrightyear{2023} 
\acmYear{2023} 
\setcopyright{acmlicensed}\acmConference[RecSys '23]{Seventeenth ACM Conference on Recommender Systems}{September 18--22, 2023}{Singapore, Singapore}
\acmBooktitle{Seventeenth ACM Conference on Recommender Systems (RecSys '23), September 18--22, 2023, Singapore, Singapore}
\acmPrice{15.00}
\acmDOI{10.1145/3604915.3608812}
\acmISBN{979-8-4007-0241-9/23/09}

\begin{document}

\title{Alleviating the Long-Tail Problem in Conversational Recommender Systems}


\author{Zhipeng Zhao}
\authornote{Both authors contributed equally to this research.}
\authornote{Contribution during internship at RUCAIBox.}
\email{zpzhao.zzp@gmail.com}
\orcid{0000-0002-1701-0286}
\affiliation{%
  \institution{School of Computer Science and Technology, Beijing Institute of Technology}
  \country{China}
}

\author{Kun Zhou}
\authornotemark[1]
\email{francis_kun_zhou@163.com}
\affiliation{%
  \institution{School of Information, Renmin University of China}
  \country{China}
}

\author{Xiaolei Wang}
\email{wxl1999@foxmail.com}
\affiliation{%
  \institution{Gaoling School of Artificial Intelligence, Renmin University of China}
  \country{China}
}

\author{Wayne Xin Zhao}
\authornote{Corresponding author}
\email{batmanfly@gmail.com}
\affiliation{%
  \institution{Gaoling School of Artificial Intelligence, Renmin University of China}
  \country{China}
}

\author{Fan	Pan}
\email{panfan3@huawei.com}
\affiliation{%
  \institution{Poisson Lab, Huawei}
  \country{China}
}
\author{Zhao Cao}
\email{caozhao1@huawei.com	}
\affiliation{%
  \institution{Poisson Lab, Huawei}
  \country{China}
}

\author{Ji-Rong	Wen}
\email{jrwen@ruc.edu.cn	}
\affiliation{%
  \institution{Gaoling School of Artificial Intelligence, Renmin University of China}
  \country{China}
}

\renewcommand{\shortauthors}{Zhao and Zhou, et al.}

\begin{abstract}
Conversational recommender systems (CRS) aim to provide the recommendation service via natural language conversations.
To develop an effective CRS, high-quality CRS datasets are very crucial.
However, existing CRS datasets suffer from the long-tail issue, \ie  a large proportion of items are rarely (or even never) mentioned in the conversations, which are called long-tail items.
As a result, the CRSs trained on these datasets  tend to recommend  frequent items,  
and the diversity of the recommended items would be largely reduced, making users easier to get bored.

To address this issue, this paper presents \textbf{LOT-CRS}, a novel framework that focuses on simulating and utilizing a balanced CRS dataset (\ie covering all the items evenly) for improving \textbf{LO}ng-\textbf{T}ail recommendation performance of CRSs.
In our approach, we design two pre-training tasks to enhance the understanding of simulated conversation for long-tail items, and adopt retrieval-augmented fine-tuning with label smoothness strategy to further improve the recommendation of long-tail items.
Extensive experiments on two public CRS datasets have demonstrated the effectiveness and extensibility of our approach, especially on long-tail recommendation.
Our code is publicly available at the link: \url{https://github.com/Oran-Ac/LOT-CRS}.
\end{abstract}

\ccsdesc[500]{Information systems~Recommender systems}

\keywords{Conversational Recommender System; Pre-trained Language Model; Long-tail Problem}

\maketitle

\input{section/introduction.tex}

\input{section/relatedwork.tex}
\input{section/preliminaries.tex}

\input{section/approach.tex}

\input{section/experiment.tex}

\section{Conclusion}
In this paper, we proposed a novel framework \textbf{LOT-CRS}, to enhance long-tail recommendation for CRSs.
First, we simulated a balanced CRS dataset  with publicly collected item attributes to resolve the data scarcity problem.
Then, based on the PLM backbone, we designed two pre-training tasks using the simulated data, namely domain-adaptive masked prediction and contrastive context alignment, to enhance the understanding of conversation context.
Finally, we adopt retrieval-augmented learning and label smoothness strategies to improve the recommendation of long-tail items using the simulated dataset.
Our approach is general to various PLM-based CRSs for improving long-tail recommendation.
Extensive experiments have shown that our approach can boost the performance of several competitive PLM-base CRS methods in terms of long-tail recommendation.

\begin{acks}
This work was partially supported by National Natural Science Foundation of China under Grant No. 62222215, Beijing Natural Science Foundation under Grant No. 4222027, fund for building world-class universities (disciplines) of Renmin University of China, and Beijing Outstanding Young Scientist Program under Grant No. BJJWZYJH012019100020098.  Xin Zhao is the corresponding author.
\end{acks}

\bibliographystyle{ACM-Reference-Format}

\end{document}

%% file: section/introduction.tex
\section{Introduction}

Nowadays, conversational intelligence has revolutionized the way that a user interacts with  online information systems. 
As a typical application of conversational intelligence, conversational recommender system (CRS)~\cite{christakopoulou2016towards,li2018towards,zhou2020improving} has emerged as a promising research topic, as it can provide the recommendation service via natural language conversations. 
Basically speaking, a CRS needs to fulfill two major functions, \ie recommending items that a user might be interested in and generating informative responses for replying to a user. 
As most existing CRSs are data hungry, their performance often relies on the availability of high-quality CRS datasets. 

Existing CRS datasets can be categorized into two types.
The first type of datasets is constructed based on real-world user-item interaction  data~\cite{sun2018conversational,lei2020estimation,zhou2020leveraging}, where the conversations are mainly simulated based on hand-crafted templates.
The other type of datasets is annotated via human chit-chats~\cite{li2018towards,hayati2020inspired,zhou2020towards}, where the annotators are asked to play the roles of the seeker or recommender and converse with each other to accomplish the goal of conversational recommendation.
These public  datasets largely advance the research on CRSs, and  a number of approaches have been developed in recent years~\cite{lei2020estimation,zhou2020improving,zhou2022c2,wang2022towards}.

\begin{figure}[t]
    \centering
    	\includegraphics[width=0.9\linewidth]{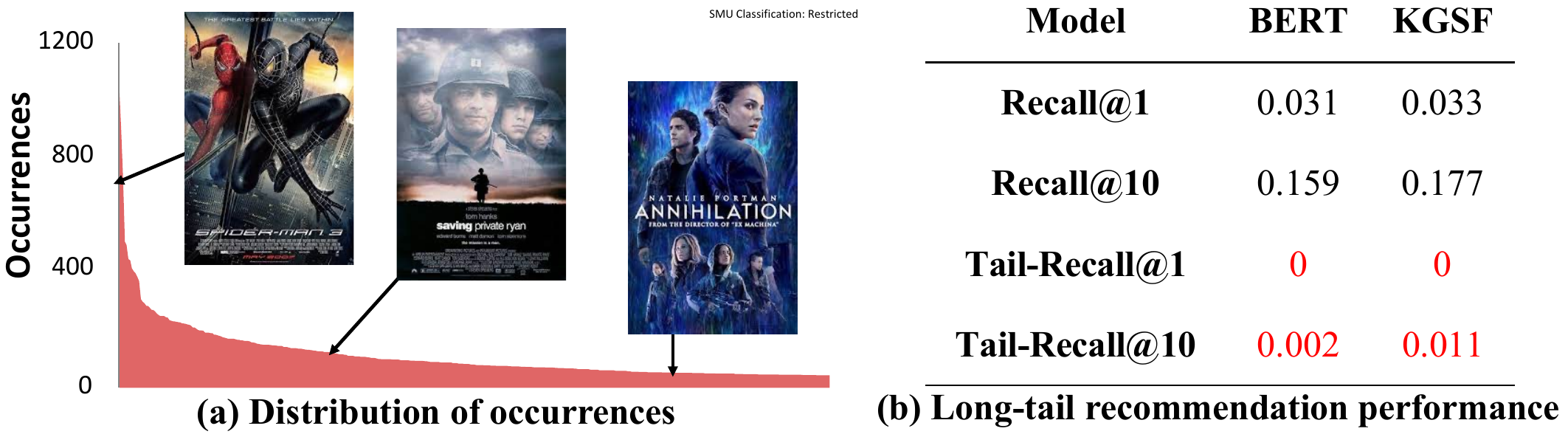}
    \caption{
        The illustration of the long-tail distribution of items in ReDial dataset~\cite{li2018towards}, and the  performance of KGSF~\cite{zhou2020improving} and BERT~\cite{devlin2019bert} on long-tail recommendation.
    }
    \label{fig:intro}
\end{figure}

However, we observe a notable  \emph{long-tail} phenomenon in existing CRS datasets: a large proportion of  items are mentioned very few times or even do not occur at all, in the conversations (see Figure~\ref{fig:intro}(a)), which are called \emph{long-tail items}\footnote{Note that, in this paper,  long-tail items refer to those infrequent items in CRS datasets, instead of  general user-item interaction datasets (\eg MovieLens~\cite{harper2015movielens}).}. In this case, the learned CRSs will be significantly affected by the data bias from the CRS dataset, \eg exposure bias~\cite{wang2020exposure,schmidt2019generalization} that a CRS tends to recommend those which are frequent in training data. To examine the capacities of existing CRSs on long-tail recommendation, we evaluate two representative CRS methods on the commonly used ReDial dataset, and report the performance with the metric \emph{Tail-Recall@$k$} (specific to long-tail items that have a frequency smaller than 4). As shown in Figure~\ref{fig:intro}(b), the performance for long-tail recommendation is indeed very poor, indicating that these models cannot effectively recommend long-tail items.

Indeed, the long-tail phenomenon is ubiquitous in real-world applications~\cite{brynjolfsson2006niches,wu2008information,park2008long}, reflecting a fundamental distribution law: only a small fraction of items receive significant attention. It has been shown that the long-tail problem will reduce the diversity of the recommended items, making users easier to get bored in recommender systems~\cite{yin2012challenging,oestreicher2012recommendation,park2008long}. 
To cope with this issue, existing works mainly adopt resampling~\cite{batuwita2010efficient,he2008adasyn,estabrooks2004multiple}, instance weighting~\cite{beutel2017beyond,cui2019class} and regularization~\cite{li2020overcoming} methods that enforce the model to focus more on the learning of long-tail items. However, these methods are not directly suited for solving the long-tail issue in CRSs, due to two major challenges. 
With highly skewed CRS datasets, it is difficult to (1) fully capture the necessary evidence to understand the conversation contexts (\eg lacking attribute information about long-tail items) and (2) accurately infer user preference (\eg lacking sufficient training signals for long-tail items). 
Existing solutions cannot effectively address the data scarcity problems in \emph{context understanding} and \emph{preference learning} of CRSs.  
Therefore, it needs a more principled approach to addressing the long-tail issue in CRSs.

To this end, in this paper, we propose a novel framework to solve the long-tail recommendation problem in CRSs. 
The key idea is to simulate a balanced CRS dataset (\ie covering all the items evenly) with publicly collected item attributes, and utilize it to resolve the data scarcity problems in context understanding and preference learning for CRSs. 
By collecting item attributes and building dialog threads, we can simulate sufficient high-quality and balanced pseudo conversations.
Based on these conversations, we design both  pre-training and fine-tuning strategies for enhancing long-tail recommendation. Specifically, we design two new pre-training tasks based on the simulated data, namely  domain-adaptive masked prediction (recovering  long-tail items and their associated attributes) and contrastive context alignment (aligning conversation contexts with the same target item), to enhance the understanding of conversation context. For fine-tuning, we adopt a retrieval-augmented learning strategy to enrich the conversation context with simulated conversations, and further employ label smoothness with a CRS model trained on simulated conversations to promote long-tail items in recommendation. Our approach can effectively improve the performance of long-tail recommendation, and is generally applicable to various PLM-based CRSs.

To the best of our knowledge, it is the first attempt to alleviate the long-tail problem in conversational recommender systems. 
We present a general framework with both pre-training and fine-tuning strategies based on a simulated CRS dataset for enhancing context understanding and preference learning in CRSs.  
To verify the effectiveness and extensibility of our approach, we conduct extensive experiments  on two CRS datasets and the results show that our framework can significantly boost the performance of existing PLM-based baselines in terms of long-tail recommendation.

%% file: section/relatedwork.tex
\section{Related Work}
Our work is related to the following two directions, namely conversational recommendation and long-tail problem.

\subsection{Conversational Recommendation}

Conversational recommender systems~(CRSs)~\cite{christakopoulou2016towards,sun2018conversational,gao2021advances} have shown great success in recent years. CRS can elicit the user's explicit preferences through natural language conversations, which are capable of accurately capturing the user's current preferences.
A category of CRS methods interact with users via pre-defined intent slots or item attributes~\cite{christakopoulou2016towards,sun2018conversational,hu2022learning,li2022user}, and aim to accomplish the recommendation task within as few turns as possible.
Multi-armed bandit algorithms~\cite{christakopoulou2016towards,xie2021comparison} and reinforcement learning~\cite{sun2018conversational} have been utilized in these methods to optimize the interaction strategy.
However, as these methods mostly rely on pre-defined intents or attributes with templates to generate responses, their usage in various scenarios is largely limited. 
Another category of CRS works relies  on natural language to converse with users and focuses on generating both accurate recommendations and human-like responses~\cite{li2018towards,hayati2020inspired,ren2022variational,zou2022improving}.
They usually devise a recommendation module and a conversation module to implement the two functions, respectively.

In fact, both the two types of CRSs require sufficient data to train the model parameters, and such training data usually derives from real-world user-item interaction records or human-annotated dialogues.
However, due to the unbalanced nature of the real-world interactions and conversations~\cite{yin2012challenging,oestreicher2012recommendation,falke2020leveraging,lin2022quantifying}, the training data would be in the long-tail distribution where a majority of items are rarely mentioned or recommended in the conversations.
As a result, these long-tail items would be hard to use by CRSs, even if they may accurately meet the niche tastes of users.
Our work focuses on this problem, and proposes a general framework to enhance the learning and recommendation of the long-tail items.

\subsection{Long-Tail Problem}
The long-tail distributions are very common in real-world information retrieval datasets~\cite{brynjolfsson2006niches,wu2008information,park2008long}, where a small fraction of popular items (\eg products, shops or documents) receive most of the user attention while most items (so-called long-tail items) only have very little feedback.
The long-tail problem refers to a set of possible problems caused by training on these datasets with long-tail distributions~\cite{yin2012challenging,oestreicher2012recommendation,park2008long}.
For example, the models trained on these datasets would easily overfit such a small fraction of items and are hard to retrieve the long-tail items.
Besides, deploying such models in real-world scenarios may further cause the ``rich get richer'' problem.
In this work, we investigate the long-tail problem of CRS, where a number of long-tail items are rarely mentioned or recommended in the conversations.

To reduce the influence of the long-tail problem, a commonly used way is the resampling method that adds redundant examples about the long-tail items to balance the distribution of the training dataset~\cite{batuwita2010efficient,he2008adasyn,estabrooks2004multiple}.
However, the resampling method easily causes overfitting on the resampled items, leading to sub-optimal performance on other items.
Another widely used way is to weight or regularize the loss of training examples, which can enforce the model to focus more on learning the long-tail items~\cite{beutel2017beyond,cui2019class,li2020overcoming}. 
Besides, there are also several works that rely on other methods to alleviate the long-tail problem, \eg transfer learning~\cite{zhang2021model}, graph-based information propagation~\cite{li2019long} and active learning~\cite{geifman2017deep}.
In this work, we first construct a simulated balanced CRS dataset, and then devise pre-training and fine-tuning strategies to enhance the understanding and learning of long-tail items based on it.

%% file: section/preliminaries.tex
\section{Problem Statement}
\label{sec-problem}
Following previous works~\cite{li2018towards,chen2019towards,zhou2020improving}, we study the task of \emph{conversational recommendation} in a recommendation-oriented conversation scenario. When a target user $u$ enters into the conversational recommender system (CRS), the conversation between user $u$ and CRS starts. At each turn, 
the CRS  asks clarified questions about the user's  preference  or intents, and the user  responds to the queries or propose specific information needs. 
The whole conversation is made through natural language utterances.
Formally, a conversation can be represented as a list of text utterances denoted as $C=\{s_t\}_{t=1}^n$, where an utterance $s_t=\{w_j\}_{j=1}^m$ at the $t$-th turn consists of a sequence of words. 
As the conversation goes on, the conversation utterances are aggregated (called \emph{conversation history} or \emph{conversation context}), and the CRS estimates the user preference over the item set (denoted as $\mathcal{I}$) based on the conversation history.

Based on the above background, the task of conversational recommendation can be described as follows. 
At the $n$-th turn, given the conversation history $C=\{s_t\}_{t=1}^{n-1}$ and the item set $\mathcal{I}$, the system should fulfill two major goals~\cite{li2018towards,hayati2020inspired,zhou2020towards}: (1) recommending a list of candidate items $\mathcal{I}_n$ (a subset from $\mathcal{I}$) and (2) generating a text utterance $R=s_{n}$ that mentions the recommended items to the user.

Unlike previous studies~\cite{lu2021revcore,zhou2020improving,wang2022towards}, we do not simply consider the overall recommendation accuracy as the only performance metric, but also emphasize that \emph{long-tail items} (with infrequent occurrences in CRS datasets) should be recommended appropriately.

%% file: section/approach.tex
\section{Approach}

In this section, we present the proposed framework \textbf{LOT-CRS} to alleviate the issue of long-tail item recommendation in CRSs.
Our framework is general to various CRS methods that rely on pre-trained language models~(PLMs) to solve recommendation and conversation tasks (Section~\ref{sec-backbone}).
In our approach, we first construct a simulated dataset that balances the frequencies of items in the conversation corpus (Section~\ref{sec-data}).
Then, we design specific pre-training (Section~\ref{sec-pretrain}) and fine-tuning strategies (Section~\ref{sec-ft}) to leverage such a simulated dataset for improving long-tail item recommendation.

\subsection{Backbone: PLM-based CRS methods}
\label{sec-backbone}

Recently, based on the powerful Transformer  architecture, pre-trained language models~(PLMs)~\cite{devlin2019bert,lewis2020bart,liu2019roberta} and large language models~(LLMs)~\cite{zhao2023survey} have shown the superiority in text and conversation understanding tasks~\cite{wu2020tod,penha2020does,zhang-etal-2022-et5}, and several CRS methods~\cite{wang2022towards,wang2021finetuning,zou2022improving}  leverage PLMs as the backbone for modeling the conversation context. 
Given the conversation context, these methods concatenate the contained utterances into a long sequence, and leverage a PLM (\eg BERT~\cite{devlin2019bert}) to encode it into contextualized representations.
Then, for the recommendation task, existing works mainly utilize the contextualized representation of the $\textsc{[Cls]}$ token as the user representation; for the conversation task, these representations are fed into a text decoder to generate proper responses.
Further, several studies~\cite{wang2022towards} also incorporate prompts into the PLM for  enriching the contextual information, \eg the knowledge graph (KG) embeddings of mentioned entities.

In this work, we aim to improve PLM-based CRSs for long-tail recommendation.
Specially, as the backbone, we consider  both classic PLM (\ie  BERT~\cite{devlin2019bert} and BART~\cite{lewis2020bart}) and recent PLM  (\ie UniCRS~\cite{wang2022towards}). 
As a general framework,  we continually pre-train these  underlying PLMs  with the simulated CRS dataset for enhancing  the understanding of conversation context on long-tail items. 
Besides, we also propose to retrieve useful evidence from the simulated CRS dataset and integrate them into the PLM during fine-tuning for fulfilling the recommendation and conversation task.  

Our approach can be considered as an improved training approach that is specially designed for long-tail recommendation, which can be generally applied to various PLM-based CRSs.

\subsection{Simulating a Balanced CRS Dataset}
\label{sec-data}

\begin{figure}
	\includegraphics[width=\linewidth]{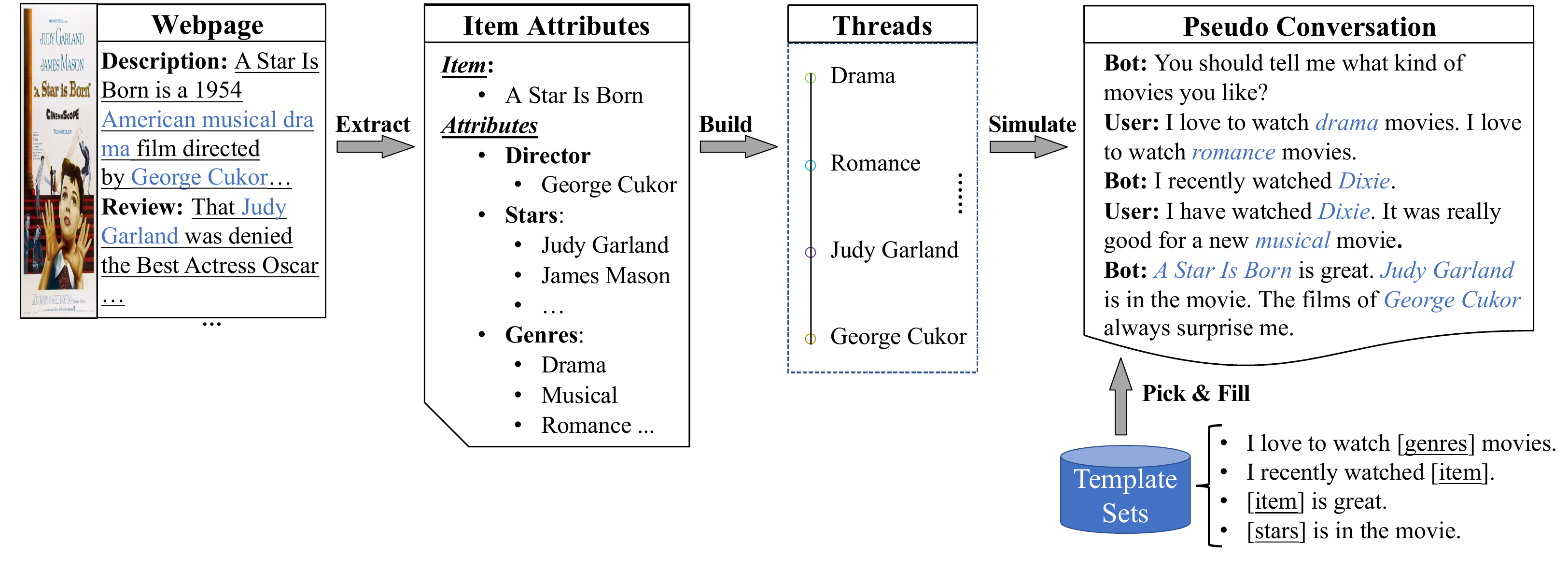}
	\centering
	\caption{
	    An illustration of our proposed pipeline to simulate an conversation example.
	    We first collect the attributes about the item from the webpage and then build the conversation thread from a random attribute to the item. Finally, we simulate the conversation based on the thread with human-written templates. 
    }
	\label{fig:data-construction}
\end{figure}
For CRSs, it is important to construct a high-quality conversation dataset for training the underlying conversational recommendation model. As we discussed before, existing datasets seldom consider long-tail items in dataset construction, which are more difficult to annotate due to lack of interaction experiences from human annotators. As the solution, our idea is to simulate a balanced CRS dataset, where all the items correspond to similar frequencies in simulated conversations.
The overview of our proposed data simulation pipeline is shown in Figure~\ref{fig:data-construction}.

\paratitle{Collecting Item Attributes.}
In order to enrich the semantics of long-tail items,
we collect a large amount of side information about items, \eg item descriptions and reviews. 
Since CRS datasets are often focused on item attributes or features,  we mainly consider  extracting  item attributes from external data resources. Specifically, we first crawl the descriptions and reviews of items from the online platform, and then extract descriptive keywords from them as the attributes of an item.  In order to identify important words, we utilize the widely used TF-IDF term weighting method to select the top-$k$ words as item attributes, in which misspelled words are removed based on the cambridge dictionary\footnote{https://dictionary.cambridge.org/}.

\paratitle{Building Conversation Threads.}
For simulating high-quality conversations about item recommendation, we build the conversation thread based on the item attributes.
In a natural real-world salesman-customer dialogue, with the conversation going on, the customer will mention more attributes about her/his preferred item, until the salesman successfully recommends it.
In this way, the mentioned attributes and items construct the thread that depicts the key information in the conversation.
To build it, we first select a common attribute (\eg type) about the item as the starting point, and then randomly add new attributes about it into the thread one-by-one.
Once the attributes in the thread can locate the target item, it will be added into the thread as the recommended item and the process will be finished.

\paratitle{Pseudo Conversations Simulation.}
Following previous simulation methods~\cite{jung2009data,zhang2020evaluating,erbacher2022state}, we extract the high-frequency conversation patterns of seeking suggestions and giving recommendations to create response templates for the user and recommender, respectively. A sample example can be given as follows. \emph{User: Can you recommend me a movie about [\underline{X}]? Recommender: I recommend the movie [\underline{Y}].} 
To simulate a complete conversation, the major step is to fill in the slots with the mentioned attributes and items from the conversation thread.
To ensure the relevance and consistency, we design several rules to ensure that (1) the attributes of co-occurring items within a conversation should have an overlap, and (2) the mentioned attribute at the $t$-th utterance should be associated with the items at the $(t-1)$-th and $(t+1)$-th utterances.
Besides, we devise heuristic  rules to control the frequencies of the filled items. 
In this way, we can construct the simulated conversations for long-tail items, in which all the items are mentioned and recommended in similar frequencies in our simulation dataset. 
Note that the simulated dataset is not aimed at improving the language ability in conversation, but mainly focuses on enriching the attribute contexts of long-tail items for recommendation. Such a simulation approach is simple and efficient, which also leads to very good performance for improving CRSs in our empirical evaluation (Section~\ref{rec-eval}). 
We leave more principled data construction approaches~\cite{li2020guided} in future work.  

To leverage the simulated conversations, we further design specific pre-training and fine-tuning strategies, which are detailed in Section~\ref{sec-pretrain} and Section~\ref{sec-ft}, respectively.

\subsection{Pre-Training on Balanced Simulation Dataset}
\label{sec-pretrain}
\begin{figure}[t]
	\includegraphics[width=\linewidth]{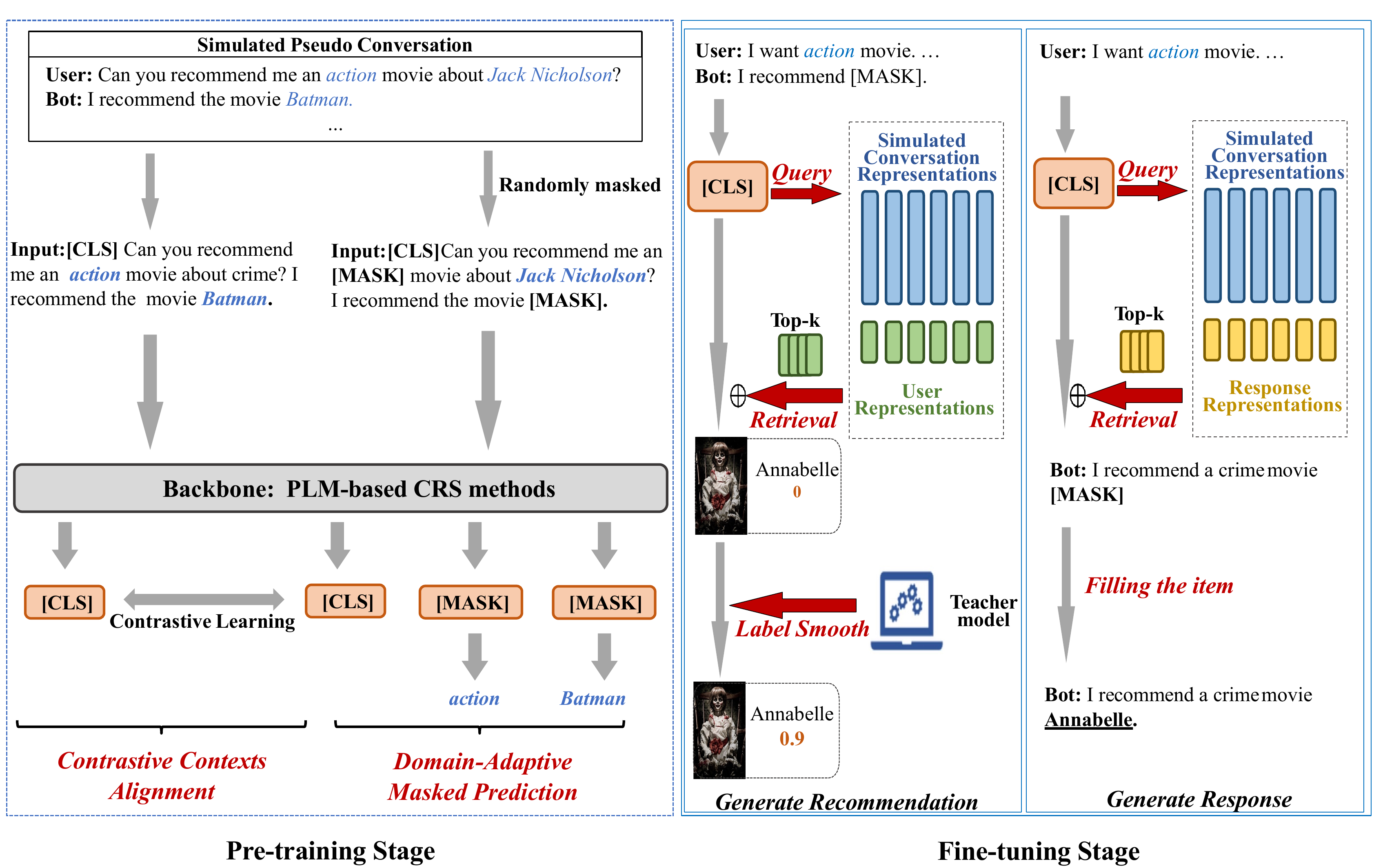}
	\centering
	\caption{We design specific pre-training and fine-tuning strategies to fully utilize the simulated balanced CRS dataset. In pre-training stage, we propose the \emph{Domain-Adaptive Masked Prediction} and \emph{Contrastive Contexts Alignment} tasks to pre-train the CRS model. In fine-tuning stage, we retrieve the top-$k$ user and response representations from the simulated dataset to improve the recommendation and response generation tasks, respectively.}
	\label{fig:approach}
\end{figure}

The simulated dataset contains important evidence about long-tail items in the conversation contexts. We design two pre-training tasks for our backbone to leverage the simulated  conversations, namely \emph{domain-adaptive masked prediction~(DMP)}, and \emph{contrastive contexts alignment~(CCA)}. The former aims to inject useful attribute information into the underlying PLMs, and the latter aims to enhance the context understanding  of the target item.

\subsubsection{Domain-Adaptive Masked Prediction}
Since the underlying PLM (\eg BERT) is pre-trained on general corpus, it is usually incapable in  conversational recommendation, especially for long-tail items. 
Inspired by the masked language model task~\cite{devlin2019bert}, we  construct pre-training tasks with 
\emph{masked item and attribute prediction} on the simulated dataset to adapt the backbone model for CRSs.

Concretely, given a simulated conversation, we randomly mask part of the contained item and attribute tokens, and then utilize the output contextualized representations of these masked tokens from the backbone model to recover them. 
Note that not all the words are considered for masked prediction, since item- and attribute-related words are more important to consider in our task. Such a selective masking strategy can alleviate the potential issue from template-based simulations, where manually generated templates are used to create the conversations. 

Thus, the loss function of this pre-training task can be defined as: 
\begin{equation}
\label{eq-mlm}
    L_{DMP} =\sum_{w_i \in W_{mask}}-\log{p(w_i | \hat{C})}
\end{equation}
where $W_{mask}$ and $\hat{C}$ denote the masked tokens and the corrupted conversation respectively, and $p(w_i | \hat{C})$ denotes the prediction probability of the masked token $w_i$.

Compared with the traditional masked language modeling (considering all the tokens), the DMP pre-training task focuses on capturing attribute- or item-level evidence from simulated conversations, in order to better understand and model long-tail items.

\subsubsection{Contrastive Contexts Alignment}
Besides the token-level pre-training, we further consider a conversation-level pre-training task tailored for conversational recommendation. For CRSs, it is key to model the semantic relatedness  between the conversation contexts and the target item, \ie the underlying model should be able to accurately recommend the target item based on the conversation contexts. For this purpose, we propose the \emph{contrastive contexts alignment} pre-training task. 

At each time, we sample two different conversations with the same target item being recommended. In this case, the contextualized representations of the two conversations should be highly similar, 
since they lead to the same recommendation (even with different recommendation reasons or conversation threads). Specifically, we utilize the encoder in the backbone model to obtain the contextualized representations (via the ``\textsc{[Cls]}'' token) of the two conversations, denoted as $\langle \bm{h}_{1}, \bm{h}_{2} \rangle$.  
To align the semantics of the two representations, we employ  contrastive learning to pull close the related pair while pulling apart other unrelated pairs:
\begin{eqnarray}
    L_{CCA}(\bm{h}_1, \bm{h}_2) = \log\frac{\exp(\text{sim}(\bm{h}_1,\bm{h}_{2})/\tau)}{\sum_{\bm{h}^{-}_j\in\left\{\bm{h}^-\right\}}{\exp(\text{sim}(\bm{h}_1,\bm{h}^{-}_{j})/\tau)}}
\label{eq-cl}
\end{eqnarray}
where $\tau$ is a temperature hyperparameter, $\text{sim}(\bm{h}_1, \bm{h}_2)$ is the cosine similarity, and $\left\{\bm{h}^-\right\}$ is the set of negative representations. In our work, we adopt the in-batch sampling strategy to construct the negative samples, where the conversations from the same mini-batch are considered as negatives. 

For long-tail items, as it is difficult to directly obtain the human-generated conversations for training the CRS, this pre-training task helps enhance the modeling of conversation context in order to better recommend long-tail items. It can effectively capture the semantic relatedness  between the conversation context and the target items. As will be shown next, it is also useful for the proposed retrieval-augmented fine-tuning strategy.

\subsection{Retrieval-Augmented Fine-Tuning}
\label{sec-ft}
Previously, we use simulated conversations (\emph{sufficient} but \emph{artificial}) for pre-training the backbone model. In this part, we further employ the human-annotated CRS dataset for fine-tuning our approach. 
However, the real CRS dataset is usually \emph{limited} or \emph{sparse} due to the huge labeling efforts, lacking sufficient coverage of all the items or necessary contexts in the conversations, which makes the fine-tuning less effective or even over-fitting. 

To address this issue, we employ the retrieval-augmented learning idea~\cite{zhu2019retrieval,guu2020retrieval,lewis2020retrieval} to effectively fine-tune our approach for CRSs, where we enrich the contexts by retrieving from the sufficient simulated conversations. 
For the recommendation task, we retrieve similar user representations for enriching the preference representation of the target user; for the conversation task, we retrieve useful utterances to help generate more informative response. Next, we present each part in detail.

\subsubsection{Fine-Tuning for Recommendation}
\label{sec-ft-rec}
The recommendation subtask aims to recommend suitable items that a user might be interested in.
It needs to learn an effective user preference representation based on the conversation context.
For this purpose, we first generate the retrieval-enhanced user representation, and then fine-tune it with a label smoothness strategy focusing on long-tail items.

\paratitle{Retrieval-Enhanced User Representation.}
As described in Section~\ref{sec-backbone}, we take the representation of the \textsc{[Cls]} token in the conversation context as the user representation $\bm{u}$.
Since the context information is often sparse in real conversations (especially for long-tail items), we propose to enrich the user representations by retrieving similar conversations from the simulated data.  
For each simulated conversation, we adopt the same way to obtain the \textsc{[Cls]} representation, and utilize the representation of the target user $u$ (\ie $\bm{u}$) to recall the top-$k$ most similar representations, denoted as 
$[\bm{u}^{'}_{1}, \cdots, \bm{u}^{'}_{k}]$.
This retrieval process can be efficiently implemented by constructing an offline dense embedding index  based on approximate nearest neighbor search algorithms~\cite{johnson2019billion}.
Furthermore, we employ a cross-attention mechanism to  obtain the retrieval-enhanced user representation $\widetilde{\bm{u}}$:

\begin{align}
 \widetilde{\bm{u}} &=\bm{u}+\gamma\cdot\sum_{j=1}^k \alpha_j\cdot \bm{u}^{'}_j,\\
 \alpha_{j} &= \frac{\exp(\bm{u}^\top\bm{W}_{1}\bm{u}^{'}_j)}{\sum^k_{j=1}\exp(\bm{u}^\top \bm{W}_{1}\bm{u}^{'}_j)},
\end{align}
where $\bm{W}_{1}$ is a learnable matrix, $\alpha_j$ is the attention weight of the retrieved user representation $\bm{u}^{'}_j$, and $\gamma$ is a hyper-parameter controlling  the influence of  retrieved user representations.
Based on it, we can compute the preference probability of 
item $i$ for user $u$ as:
\begin{align}
    \label{eq:rec-ent}
    \text{Pr}(i) & =\text{softmax}(\widetilde{\bm{u}}^\top \cdot\bm{e}_i),
\end{align}
where $\bm{e}_i$ is the embedding of item $i$.

\paratitle{Label Smoothness for Long-Tail Items.}
To fine-tune the CRS, the cross-entropy loss has been widely used:
\begin{equation}
    \label{eq:rec-loss-p}
    L_{CE}=-\sum_{j=1}^n \sum_{i=1}^m\big[y_{j,i}\cdot\log\text{Pr}_{j}(i)+(1-y_{j,i})\cdot\log(1-\text{Pr}_{j}(i))\big],
\end{equation}
where $n$ is the number of training instances (\ie a pair of the context and a target item), $m$ is the total number of items, and $y_{j,i}$ denotes a binary ground-truth label which is equal to 1 when item $i$ is the correct label for the $j$-th training instance. However, long-tail items rarely appear in real CRS dataset, such a loss is likely to lead to the issue of \emph{exposure bias}~\cite{schmidt2019generalization,wang2020exposure}, \ie tending to predict those frequent items in the CRS dataset. To rectify this issue, we propose a label smoothing strategy to promote long-tail items. The basic idea is to pre-train a CRS model (called \emph{teacher model})\footnote{For efficiency, we adopt a pre-learned KG-based model KGSF~\cite{zhou2020improving} as the teacher model, which can be replaced by other more effective CRS methods.} based on the simulated data, then employ it to produce soft labels (denoted as $y^{'}_{j,i}$) on the real CRS dataset. Based on it, we construct a regularizer that asks the backbone model to mimic the predictions from the teacher model.

Formally, the regularizer minimizes the Kullback-Leibler Divergence between our model and teacher model as:
\begin{equation}
    \label{eq:kl}
    L_{SOFT}=-\sum_{j=1}^n\sum_{i=1}^m D_{KL}(y^{'}_{j,i}, y_{j,i}). 
\end{equation}
With this regularizer, we can enforce the model to generate higher probabilities to long-tail items.
Besides, we also add the CCA loss (Eq.~\ref{eq-cl}) to enhance the semantic connection between conversation context and target item, which ensures the effectiveness of the retrieval module, \ie highly similar contexts should correspond to same target items.
Finally, the joint loss function for the recommendation task is:
\begin{equation}
\label{eq-rec-all}
    L_{REC} = L_{CE} +\lambda_{1} \cdot L_{SOFT} +\lambda_{2} \cdot L_{CCA},
\end{equation}
where $\lambda_1$ and $\lambda_2$ are hyper-parameters to trade off the three losses.

\subsubsection{Fine-Tuning for Response Generation}
The response generation task focuses on generating fluent and informative responses to reply to users. 
As described in Section~\ref{sec-backbone}, we utilize the generative PLM in existing methods for response generation.

\paratitle{Retrieval-Enhanced Prompts}.
Given the conversation history $C$, we utilize our backbone model to encode its tokens into contextualized representations, denoted as $[\bm{h}_{w_{1}}, \cdots, \bm{h}_{w_{m}}]$.
Then, we retrieve relevant responses from the simulated datasets for helping generate high-quality responses. At this step, we consider the \textsc{[Cls]} token representations of all the simulated conversations for retrieval. With the \textsc{[Cls]} representation of current conversation as query, we can obtain  top-$k$ representations of the responses, denoted as $[\bm{r}^{'}_{1}, \cdots, \bm{r}^{'}_{k}]$.

Then, we fuse these representations and the contextualized representations of the input by cross attention:

\begin{align}
 \widetilde{\bm{h}}_{w_{1}}&=\bm{h}_{w_{1}}+\sum_{j=1}^k \beta_j\cdot \bm{r}^{'}_j,\\ 
 \beta_{j} &= \frac{\exp(\bm{h}_{w_{1}}^\top\bm{W}_{2}\bm{r}^{'}_j)}{\sum^k_{j=1}\exp(\bm{h}_{w_{1}}^\top\bm{W}_{2}\bm{r}^{'}_j)}, 
\end{align}

where $\bm{W}_{2}$ is a learnable matrix, $\beta_j$ is the attention weight of the retrieved response representation $\bm{r}^{'}_j$, and $[\widetilde{\bm{h}}_{w_{1}}, \cdots, \widetilde{\bm{h}}_{w_{m}}]$ are the retrieval-enhanced contextual representations, which will be used as the soft prompts to guide the generative PLM to generate more informative responses.

\paratitle{Response Template Generation and Filling.}
After obtaining the retrieval-enhanced prompts, we integrate them as the prefix of the conversation context in the embedding layer of the generative PLM to generate the response templates.

A naive approach is to let the PLM directly generate both the response and the recommended item. 
However, the PLM has rarely seen the recommended items during pre-training on large-scale dialog corpus, hence we replace all items in responses by a special token \textsc{[Item]}.
In this way, the prompted PLM generates either the \textsc{[Item]} or other general token at each decoding step, and the training objective is formulated as:
\begin{align}
    L_{GEN} = -\frac{1}{n}\sum_{i=1}^n \sum_{j=1}^{l_i}\log \text{Pr}(w_{i,j} \mid C_{i} ; w_{<j}),
 \label{eq:conv-loss}
\end{align}
where $l_i$ is the length of the $i$-th response, and $w_{<j}$ denotes the tokens proceeding the $j$-th position.
During inference, all the \textsc{[Item]} tokens within the generated template will be filled by the predicted items from the recommendation module.

\paratitle{Summary of Our Approach}. Our approach is developed based on a simulated CRS dataset, where it balances the distribution of the items in generated conversations. Such a dataset is mainly built on item or attribute information extracted from public resources. We first introduce two pre-training tasks, \ie  domain-adaptive masked prediction (DMP) and contrastive context alignment (CCA) tasks using Eq.~\ref{eq-mlm} and Eq.~\ref{eq-cl}, which inject knowledge information of long-tail items into underlying PLMs and enhance the understanding of conversation contexts, respectively. Furthermore, we employ a retrieve-augmented fine-tuning approach to enriching the limited context in real conversations. Specifically, we consider incorporating regularizer to promote long-tail items (Eq.~\ref{eq-rec-all}) for recommendation, and generating retrieving-enhanced prompts for generating more informative response (Eq.~\ref{eq:conv-loss}).

%% file: section/experiment.tex
\section{Experiment}

\subsection{Experimental Setup}

\paratitle{Datasets.}
To verify the effectiveness of our approach, we conduct experiments on two widely used English CRS datasets, \textsc{ReDial}~\cite{li2018towards} and \textsc{INSPIRED}~\cite{hayati2020inspired}.
\textsc{ReDial} is about movie recommendations, and is constructed through crowd-sourcing workers on Amazon Mechanical Turk (AMT).
Similar to \textsc{ReDial}, \textsc{INSPIRED} is also about movie recommendations, but with a smaller size.

The statistics of both datasets are summarized in Table~\ref{tab:datasets}.

\begin{table}[t]
    \centering
    \caption{Statistics of the datasets after preprocessing. }
    \label{tab:datasets}
    \begin{tabular}{cccccc}
        \toprule
        \textbf{Dataset} & \textbf{\#Dialogs} & \textbf{\#Utterances} & \textbf{\#Items} &  \\
        \midrule
        INSPIRED         & 1,001              & 35,811                & 1,783     \\
        ReDial           & 10,006             & 182,150               & 51,699          \\
        \bottomrule
    \end{tabular}
\end{table}

\paratitle{Baselines.}
We select several competitive methods (including both CRS models and adapted PLMs) as baselines, and evaluate them on the two subtasks, namely recommendation and conversation.
    
$\bullet$ \underline{\textbf{ReDial}}~\cite{li2018towards} incorporates a conversation module based on HRED~\cite{subramanian2018learning} and a recommendation module based on auto-encoder~\cite{sedhain2015autorec}.

$\bullet$ \underline{\textbf{KBRD}}~\cite{chen2019towards} utilizes an external KG to enhance the entities in the dialogue history, and adopts a self-attention based recommendation module and a Transformer-based conversation module.

$\bullet$ \underline{\textbf{KGSF}}~\cite{zhou2020improving} incorporates two KGs to enhance the representations of words and entities, and utilizes the Mutual Information Maximization method to align the semantic spaces of the two KGs.

$\bullet$ \underline{\textbf{GPT-2}}~\cite{radford2019language} is an auto-regressive PLM. We use the dialog context as input, and take the generated text as the response and the representation of the last token for recommendation.

$\bullet$  \underline{\textbf{DialoGPT}}~\cite{zhang2020dialogpt} is an auto-regressive model pre-trained on a large-scale dialogue corpus. We also adopt the generated text as the response and the last token representation for recommendation.

$\bullet$  \underline{\textbf{BERT}}~\cite{devlin2019bert} is pre-trained via the masked language model task. 
We utilize the representation of \textsc{[CLS]} token for recommendation.

$\bullet$ \underline{\textbf{BART}}~\cite{lewis2020bart} is a seq2seq model pre-trained with the denoising auto-encoding task.
We also adopt the generated text as the response and the last token representation for recommendation.

$\bullet$  \underline{\textbf{UniCRS}}~\cite{wang2022towards} is a unified model that fulfills both the recommendation and conversation tasks, using knowledge-enhanced prompt learning on a fixed PLM.

\paratitle{Evaluation Metrics.}
Following existing CRS works~\cite{li2018towards,chen2019towards,zhou2020improving}, we adopt different metrics to evaluate the recommendation and conversation task separately.
For recommendation, we use Recall@$k$ ($k$=1,10,50) to evaluate the recommendation accuracy, and also select long-tail related metrics~\cite{kim2019sequential,liu2020long}:

(1) \textbf{Coverage@$k$} ($k$=10,50) is used to measure how many different items appear across all the recommendation lists among total items, which measures the diversity of the recommended results, formulated as: 
\begin{equation}
\text{Coverage}@k=\frac{|\bigcup_{C\in \mathcal{C}}L_{k}(C)|}{|\mathcal{I}|},
\end{equation}
where $L_{k}(C)$ represents the list of top-$k$ recommended items for the conversation $C$. 

(2) \textbf{Tail-Coverage@$k$}  ($k$=10,50) is a variant of Coverage@$k$, 
which focuses on measuring the number of different long-tail items across all recommendation lists, formulated as: 
\begin{equation}
\text{Tail\_Coverage}@k=\frac{|\bigcup_{C\in \mathcal{C}}L_{k}^{(T)}(C)|}{|\mathcal{I}^{(T)}|},
\end{equation}
where superscript $(T)$ denotes the tail-items (with a frequency smaller than 4 in our corpus), and $L_{k}^{(T)}(C)$ is a subset of $L_{k}(C)$ containing long-tail items from $\mathcal{I}^{(T)}$. 

For conversation, we adopt Distinct-$n$ ($n$=2,3,4) at the word level to evaluate the diversity of the generated responses.
Following \cite{zhou2020improving}, we invite three annotators to score (ranging 0 to 2) the generated responses of our model and baselines from the aspects of \emph{Fluency} and \emph{Informativeness}. 
For all the metrics, we  report the average scores on all test examples.

\paratitle{Implementation Details.}
All baselines are implemented using CRSLab~\cite{zhou2021crslab} or the released code from original papers, and adopt the same hyper-parameters setting as authors suggested.

For the hyper-parameters about knowledge-enhanced prompts, we follow the same setting as KGSF~\cite{zhou2020improving} and UniCRS~\cite{wang2022towards}.
For LOT-CRS+BERT and LOT-CRS+BART, we select the BERT-base model~\cite{devlin2019bert} and BART-base model~\cite{lewis2020bart} as the base PLM respectively, which consists of 12 Transformer layers, and the dimension of its embeddings is 768.
In the conversation task, we utilize the DialoGPT-small model~\cite{zhang2020dialogpt} to generate the response template, and freeze all its parameters.
During training, we use AdamW~\cite{loshchilov2018decoupled} with the default parameter setting to optimize the parameters in our backbone model.
The batch size is set to 32 for the recommendation task and 16 for the conversation task, and the learning rate is 5e-5 for  pre-training, 1e-5 for the recommendation task and 1e-4 for the conversation task.

\subsection{Evaluation on Recommendation Task}
\label{rec-eval}
In this part, we evaluate our approach on recommendation task.

\begin{table*}[t]
    \small
    \centering
    \caption{
        Results on the recommendation task. The best methods in each group are marked in bold.
        Numbers marked with * indicate that the improvement is statistically significant  (t-test with p-value < 0.05). 
        R@$k$, C@$k$ and TC@$k$ are the abbreviations of Recall@$k$, Coverage@$k$ and Tail-Coverage@$k$, respectively. \emph{+Ours} means that we implement our framework on the baseline.
    }
    \label{tab:rec-table}
        \begin{tabular}{c|c|*{5}{c} |cl|cc|cl}
            \toprule
            Datasets& Metrics & Redial & KBRD & KGSF & GPT-2 & DialoGPT & BERT & +Ours &  BART & +Ours & UniCRS & +Ours  \\
            \midrule
            \multirow{7}*{\textsc{Redial}}&  R@1 & 0.023 & 0.033 & 0.035 & 0.023 & 0.030 &0.030 & \textbf{0.049}$^*$ & 0.034 & \textbf{0.048}$^*$ & 0.051 & \textbf{0.054}$^*$ \\
             ~&R@10 & 0.129 & 0.175 & 0.177 & 0.147 & 0.173 & 0.156 & \textbf{0.216}$^*$ & 0.174 & \textbf{0.224}$^*$ & 0.224 & \textbf{0.238}$^*$ \\
             ~&R@50 &0.287 & 0.343 & 0.362 & 0.327 & 0.361 & 0.357 & \textbf{0.416}$^*$ & 0.377 & \textbf{0.413}$^*$ & 0.420 & \textbf{0.424} \\
             ~&C@10 & 0.001 & 0.229 & 0.295 & 0.237 & 0.232 & 0.127 & \textbf{0.324}$^*$ & 0.138 & \textbf{0.264}$^*$ & 0.318 & \textbf{0.325}$^*$ \\
             ~ & C@50 & 0.008 & 0.432 & 0.462 & 0.483 &0.480 & 0.281 & \textbf{0.605}$^*$ & 0.314 & \textbf{0.491}$^*$ & 0.576 & \textbf{0.581}$^*$ \\
             ~ & TC@10 & 0.002 & 0.082 & 0.140 & 0.055 & 0.046 & 0.014 & \textbf{0.090}$^*$ & 0.017 & \textbf{0.046}$^*$ & 0.117 & \textbf{0.150}$^*$ \\
             ~ & TC@50 & 0.002 & 0.266 & 0.300 & 0.274 & 0.270 & 0.087 & \textbf{0.293}$^*$ & 0.117 & \textbf{0.191}$^*$ & 0.351 & \textbf{0.432}$^*$ \\
             \midrule
             \multirow{7}*{\textsc{INSPIRED}} & R@1 & 0.003 & 0.045 & 0.055 & 0.034 & 0.024 & 0.043 & \textbf{0.051}$^*$ & 0.037 & \textbf{0.055}$^*$ & 0.084 & \textbf{0.086} \\
             ~ & R@10 & 0.117 & 0.142 & 0.178 & 0.112 & 0.125 & 0.148 & \textbf{0.160}$^*$ & 0.132 & \textbf{0.148}$^*$ & 0.200 & \textbf{0.218}$^*$ \\
             ~ & R@50 & 0.285 & 0.233 & 0.256 & 0.278 & 0.247 & 0.320 & \textbf{0.324} & 0.247 & \textbf{0.285}$^*$ & 0.350 & \textbf{0.363}$^*$ \\
             ~ & C@10 & 0.007 & 0.010 & 0.012 & 0.048 & 0.024 & 0.051 & \textbf{0.122}$^*$ & 0.031 & \textbf{0.106}$^*$ & 0.339 & \textbf{0.361}$^*$ \\
             ~ & C@50 & 0.029 & 0.041 & 0.055 & 0.190 & 0.089 & 0.196 & \textbf{0.467}$^*$ & 0.118 & \textbf{0.471}$^*$ & 0.607 & \textbf{0.637}$^*$ \\
             ~ &TC@10 & 0.003 & 0.003 & 0.004 & 0.010 & 0.002 & 0.026 & \textbf{0.066}$^*$ & 0.004 & \textbf{0.052}$^*$ & 0.308 & \textbf{0.334}$^*$ \\
             ~ & TC@50 & 0.015 & 0.018 & 0.027 & 0.141 & 0.042 & 0.153 & \textbf{0.393}$^*$ & 0.067 & \textbf{0.399}$^*$ & 0.588 & \textbf{0.621}$^*$ \\

            \bottomrule
        \end{tabular}%
\end{table*}

\paratitle{Automatic Evaluation.}
Table~\ref{tab:rec-table} shows the performance of different methods on accuracy and long-tail metrics about the recommendation task.
First, the performance of KG-based CRS methods (\ie KBRD and KGSF) on recommendation accuracy metrics is competitive with PLMs (\ie GPT-2, DialoGPT, BERT and BART).
It indicates the effectiveness of external KGs, which can enrich the semantics of entities in the conversation history and alleviate the data sparsity problem.
For the metrics Coverage@$k$ and Tail-Coverage@$k$ that measure the diversity of recommended items and long-tail ones, we can see that KG-based methods also perform better.
It is because that these models can enrich the semantics of entities with neighbour nodes on KGs, which can enhance the representations of long-tail items, leading to more diverse recommendation results.
Among all the baselines, UniCRS achieves the best performance on most of metrics. UniCRS incorporates knowledge-enhanced prompts on a PLM to fuse the semantics of KG and PLM, which can improve the recommendation accuracy and enrich the long-tail items.

Finally, our approach can boost the performance of BERT, BART and UniCRS on both recommendation accuracy and long-tail related metrics.
It demonstrates  the effectiveness of our approach on improving the long-tail recommendation of PLM-based CRS methods.
Our approach can leverage a simulated CRS dataset that covers all the items evenly, and incorporates both pre-training and fine-tuning strategies  for enhancing the context understanding and preference learning with this dataset.
Such a way can effectively inject the background knowledge of items into PLMs  for accurate recommendation.
Besides, we also adopt the retrieval-augmented learning and  label smoothness for fine-tuning, which further promotes the long-tail items in recommendation.

\begin{table*}[t]
    \small
    \centering
    \caption{
        Results on the recommendation task with long-tail methods on the \textsc{ReDial} dataset. The best methods are marked in bold. 
     \emph{+CE$_{cbs}$} and \emph{+Ours} means that we implement Class-Balanced Softmax Cross-Entropy Loss and our framework on BERT, respectively. 
    }
    
    \label{tab:rec-table-longtail}
        \begin{tabular}{l|*{8}{c}}
            \toprule
            Methods  & R@1 & R@10 & R@50 & C@10 & C@50 & TC@10 & TC@50  \\
            \midrule
             BERT     & 0.030 & 0.156 & 0.357 & 0.127 & 0.281 & 0.014 & 0.087   \\
             \midrule
             BERT + Regularization  & 0.025 & 0.150 & 0.350 & 0.227 & 0.381 & 0.054 & 0.187  \\
             BERT + CE$_{cbs}$ & 0.028 & 0.146 & 0.277 & 0.220 & 0.319 & 0.069 & 0.203  \\
             BERT + Ours &\textbf{0.049}$^*$ & \textbf{0.216}$^*$ & \textbf{0.416}$^*$ & \textbf{0.324}$^*$ & \textbf{0.605}$^*$ & \textbf{0.090}$^*$ & \textbf{0.293}$^*$ & \\
             
 \bottomrule
        \end{tabular}%
\end{table*}

\paratitle{Performance Comparison with Other Long-tail Methods.} 
In this part, we compare our approach with two widely-used methods to alleviate the long-tail problems: (1) \emph{Class-Balanced Softmax Cross-Entropy Loss}~\cite{cui2019class} and (2) \emph{Regularization}~\cite{li2020overcoming}. We adopt BERT as the backbone and conduct the experiments on ReDial dataset.
As shown in Table~\ref{tab:rec-table-longtail}, all long-tail methods can boost performance of BERT on long-tail metrics but perform not better than our approach. 
Since our framework simulates a balanced CRS dataset that can be utilized for both pre-training and fine-tuning.
As a comparison, both the long-tail methods only focus on the fine-tuning process and cannot well address the intrinsic long-tail problem in the training data.

\paratitle{Ablation Study.}
Our approach consists of several components to improve the long-tail recommendation.
To verify their effectiveness, we implement our approach on the best baseline UniCRS, and conduct the ablation study on the \textsc{ReDial} dataset. 
We report the results of Recall@10 and Tail-Coverage@50.
Here, we consider removing the two pre-training tasks (\ie DMP in Eq.~\ref{eq-mlm} and CCA in Eq.~\ref{eq-cl}), the label smoothness  and retrieval-augmented strategies, respectively.
The results are shown in Figure~\ref{fig:ablation-rec}, where LS and RET denote the label smoothness and retrieval-augmented strategies, respectively.
First, we can see that removing any component or technique would lead to performance degradation, especially for Tail-Coverage@50.
It indicates that the proposed techniques in our model are useful to improve the performance in terms of long-tail recommendation.
Besides, after removing DMP, the performance significantly decreases. It indicates that such a pre-training task is more important, since it can enhance the understanding of long-tail items in conversations.

\paratitle{Hyper-parameters Analysis.}
In our approach, there are two important hyper-parameters to tune, \ie the number of retrieved user representations $k$ and the weight of the retrieved user representations $\gamma$.
In this part, we investigate the effect of each hyper-parameter in our approach.
\begin{figure*}[t]
\begin{minipage}[t]{0.49\linewidth}
    \centering
    \includegraphics[width=0.49\linewidth]{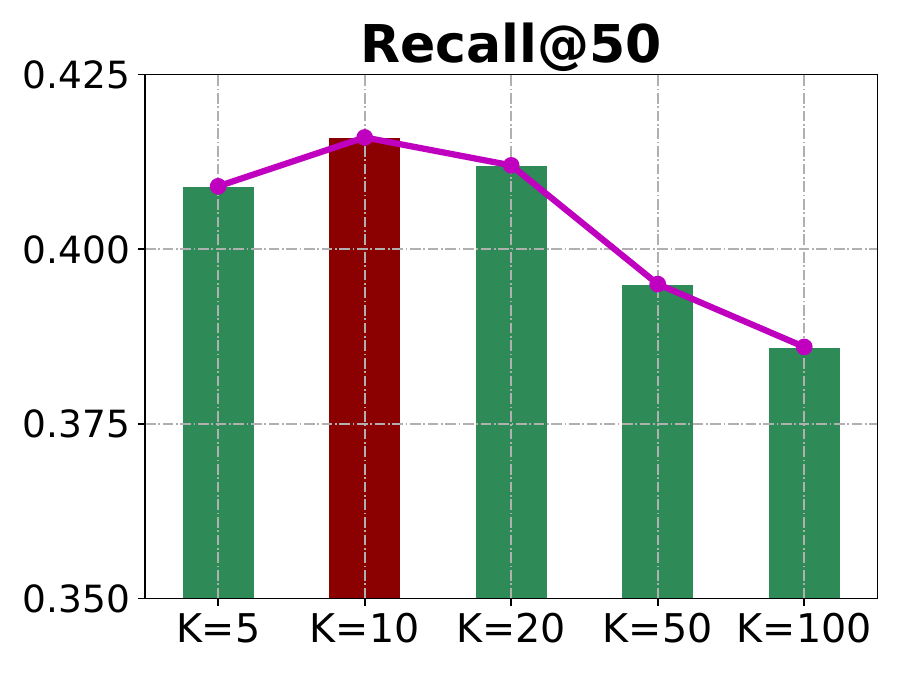}
    \includegraphics[width=0.49\linewidth]{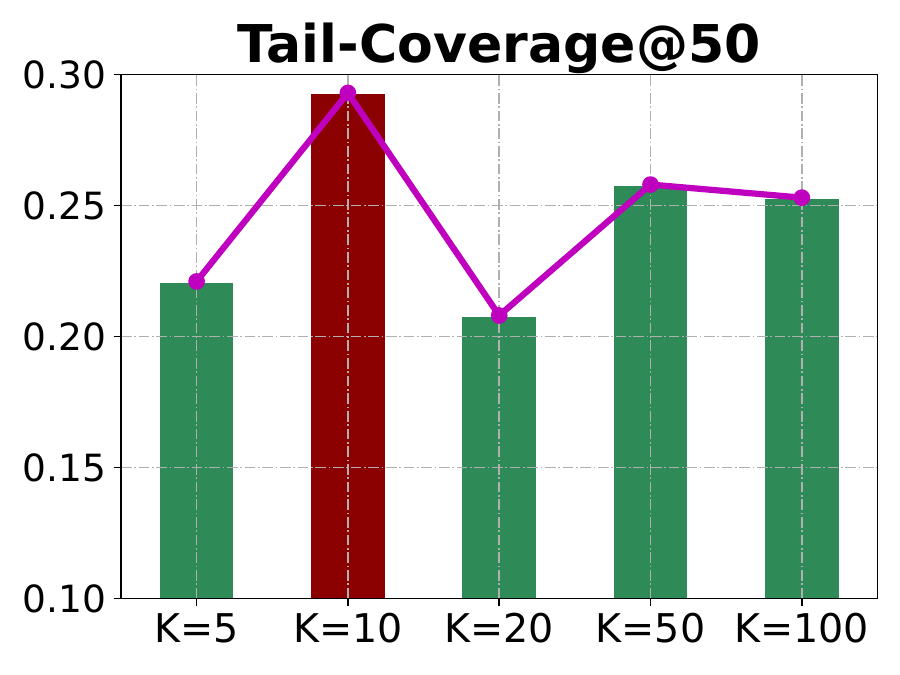}
    \caption{The recommendation performance of our approach in UniCRS on \textsc{ReDial} \emph{w.r.t.} different numbers of retrieved user representations $k$.}
    \label{fig:Hyper-parameters-k}
\end{minipage}
\hfill
\begin{minipage}[t]{0.49\linewidth}
\centering
    \includegraphics[width=0.49\linewidth]{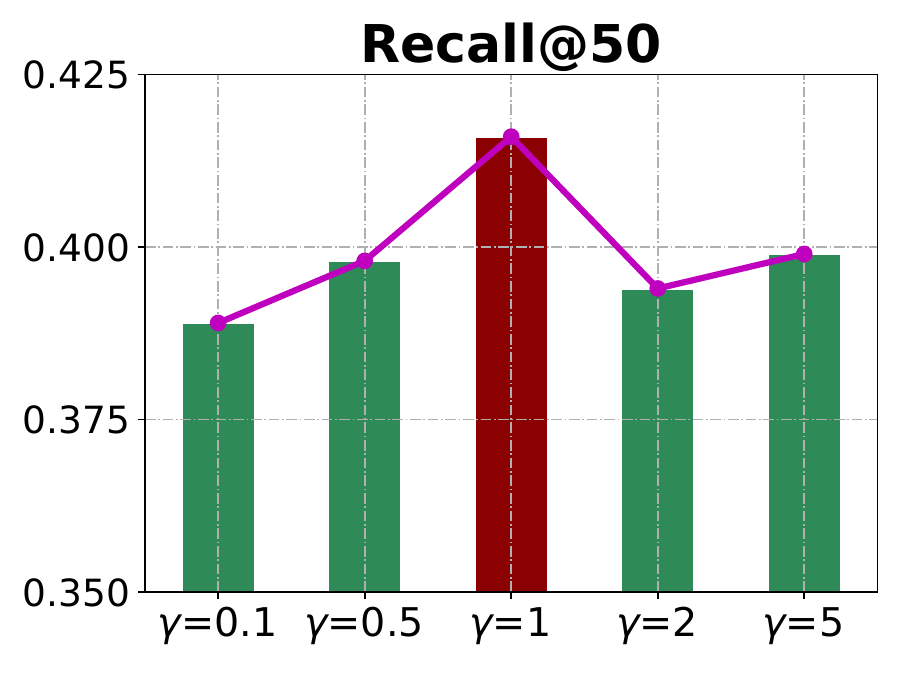}
    \includegraphics[width=0.49\linewidth]{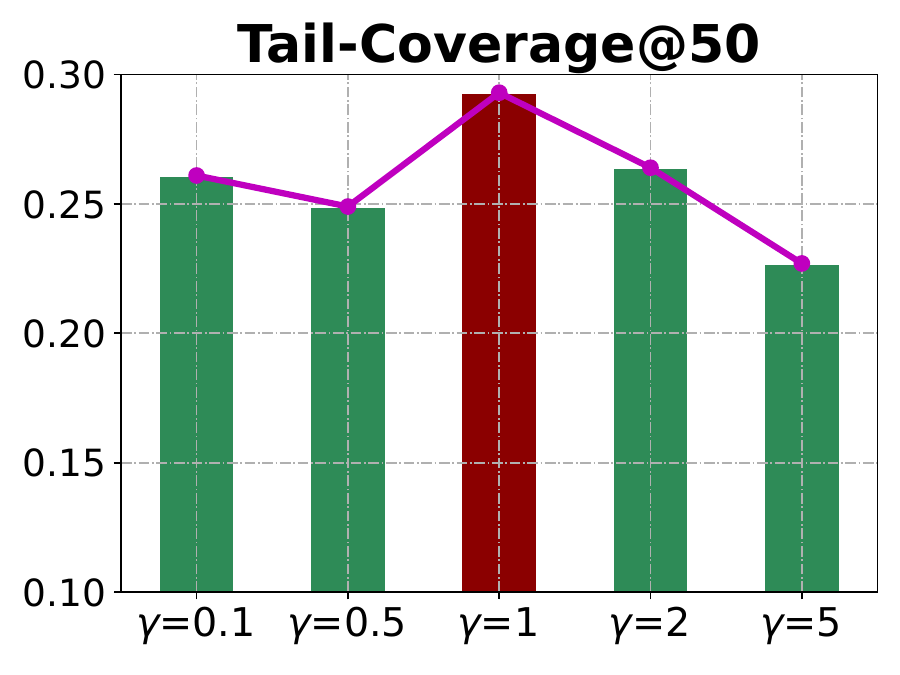}
    \caption{The recommendation performance of our approach in UniCRS on \textsc{ReDial} \emph{w.r.t.} different weights of the retrieved user representation $\gamma$.}
    \label{fig:Hyper-parameters-gamma}
\end{minipage}
\end{figure*}
We also implement our approach on UniCRS, and conduct the analysis experiments on the \textsc{ReDial} dataset. 
We report the results of Recall@10 and Tail-Coverage@50. 
As shown in Figure~\ref{fig:Hyper-parameters-k} and Figure~\ref{fig:Hyper-parameters-gamma},  too large or too small values of the two hyper-parameters would lead to the performance degradation: 
Too large $k$ and $\gamma$ may cause the final user representation to be  dominated by the retrieved results, and too small ones may reduce the influence of the retrieved representations.
We should set them to appropriate values to ensure the performance.

\subsection{Evaluation on Conversation Task}

In this part, we evaluate our approach on conversation task.

\begin{table*}[t]
    \centering
    \small
    \caption{
        Automatic evaluation results on the conversation task.
        We abbreviate Distinct-2,3,4 as Dist-2,3,4.
    }
    \label{tab:conv-table}
        \begin{tabular}{lcccccc}
            \toprule
            Datasets & \multicolumn{3}{c}{ReDial} & \multicolumn{3}{c}{INSPIRED}                                                                         \\
            \midrule
            Models   & Distinct-2                     & Distinct-3                       & Distinct-4          & Distinct-2          & Distinct-3          & Distinct-4          \\
            \midrule
            ReDial   & 0.225                      & 0.236                        & 0.228           & 0.406           & 1.226           & 2.205           \\
            KBRD     & 0.281                      & 0.379                        & 0.439           & 0.567           & 2.017           & 3.621           \\
            KGSF     & 0.302                      & 0.433                        & 0.521           & 0.608           & 2.519           & 4.929           \\
            GPT-2    & 0.354                      & 0.486                        & 0.441           & 2.347           & 3.691           & 4.568           \\
            DialoGPT & 0.476                      & 0.559                        & 0.486           & 2.408           & 3.720           & 4.560           \\
            \midrule
            BART     & 0.376                      & 0.490                        & 0.435           & 2.381           & 2.964           & 3.041           \\
            Ours+BART &\textbf{0.396}$^{*}$ & \textbf{0.546}$^{*}$ & \textbf{0.573}$^{*}$ & \textbf{2.697}$^{*}$ & \textbf{4.074}$^{*}$ & \textbf{4.811}$^{*}$ \\
            \midrule
            UniCRS   & 0.492    & 0.648 & 0.832 & 3.039 & 4.657 & 5.635 \\
            Ours+UniCRS  & \textbf{0.546}$^{*}$ & \textbf{0.910}$^{*}$ & \textbf{1.158}$^{*}$ & \textbf{4.250}$^{*}$ & \textbf{6.425}$^{*}$ & \textbf{7.475}$^{*}$ \\
            \bottomrule
        \end{tabular}%
\end{table*}

\paratitle{Automatic Evaluation.}
Table~\ref{tab:conv-table} shows the results of automatic metrics about different methods.
As we can see, PLM-based methods consistently perform better than the other CRS methods. Since they have been pre-trained with generative tasks on a large-scale corpus,  they can better adapt to the response generation task of CRS by fine-tuning. 
Among these PLMs, DialoGPT achieves the best performance, as it has been pre-trained on a large-scale dialogue corpus.
In addition, UniCRS also outperforms the other baselines, since it  leverages the prompt learning  strategy to inject related knowledge into the PLM.
Finally, our model also consistently improves the performance of BART and UniCRS, and \emph{Ours+UniCRS} achieves the best performance among all the methods.
In our approach, we not only pre-train our backbone model to learn the useful knowledge for CRS, but also retrieve useful responses from the simulated balanced CRS dataset.
In this way, we can effectively learn the background information about items, and generate more informative responses.

\paratitle{Human Evaluation.}
We conduct the human evaluation to further test the effectiveness of our method. 
We select KGSF, DialoGPT and UniCRS for comparison due to their better performance on automatic metrics, and we implement our approach on UniCRS.
Table~\ref{tab:human-table} presents the results on the \textsc{ReDial} dataset.
Among the three baseline methods, UniCRS performs the best in both metrics. UniCRS fuses the semantics from KG and PLM to model the conversation history, which is more effective than KGSF and DialoGPT that only utilize KG or PLM.
Besides, our approach with UniCRS also outperforms all the baseline models. In our approach, we simulate a  balanced CRS dataset to enrich the information about items during both pre-training and fine-tuning.
In this way, our model can effectively understand the dialogue history, and generate fluent and informative responses.

\begin{table*}[t]
     \centering
     \small
     \caption{Human evaluation results about the conversation task on the \textsc{ReDial} dataset. }
     \label{tab:human-table}
         \begin{tabular}{lccc|c}
             \toprule
             \textbf{Aspects} &    \textbf{KGSF} &    \textbf{DialoGPT} &    \textbf{UniCRS} &    \textbf{Ours+UniCRS} \\
             \midrule
             \emph{Fluency} &1.49 & 1.68 & 1.72 & \textbf{1.84} \\
             \emph{Informativeness} & 1.39 & 1.56 &1.64 & \textbf{1.75} \\
             \bottomrule
         \end{tabular}%
 \end{table*}
 
\begin{figure*}[t]
\begin{minipage}[t]{0.48\linewidth}
    \centering
    \includegraphics[width=0.49\linewidth]{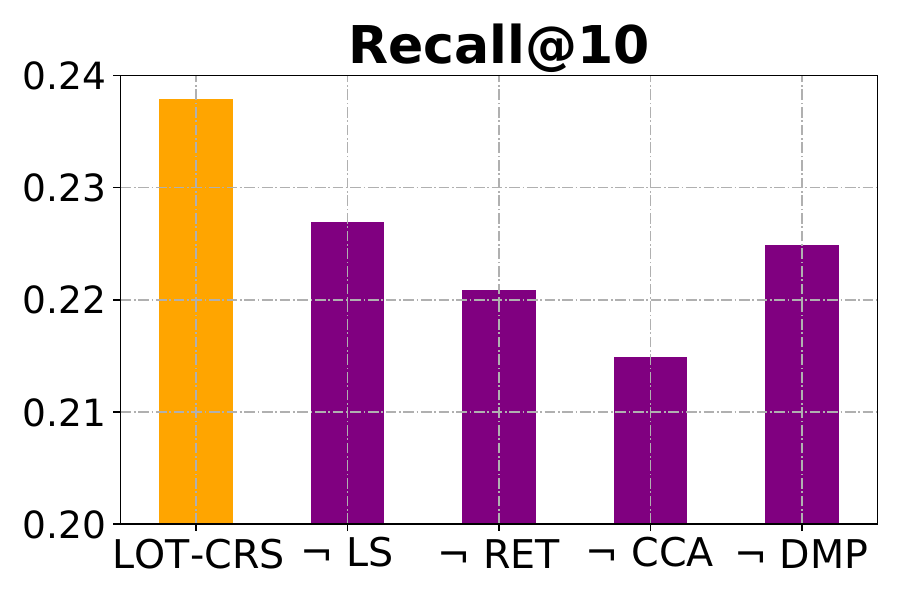}
    \includegraphics[width=0.49\linewidth]{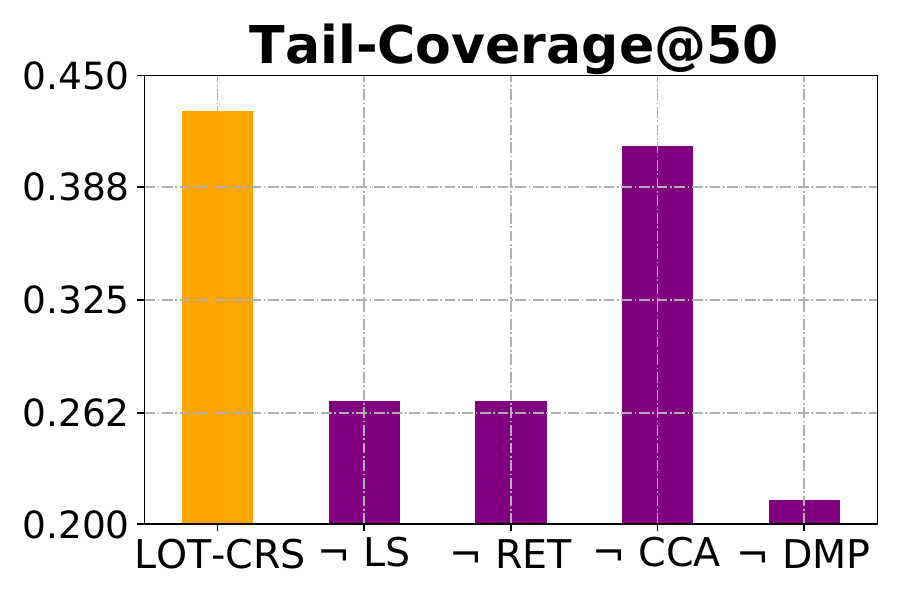}
    \caption{
        Ablation study on the \textsc{ReDial} dataset about the recommendation task using UniCRS as the backbone.
        LS denotes label smoothness, RET refers to retrieval-enhanced prompt, CCA and DMP denote the two pre-training tasks, \ie contrastive contexts alignment and domain-adaptive masked prediction, respectively.
    }
    \label{fig:ablation-rec}
\end{minipage}
\hfill
\begin{minipage}[t]{0.48\linewidth}
    \centering
    \includegraphics[width=0.49\linewidth]{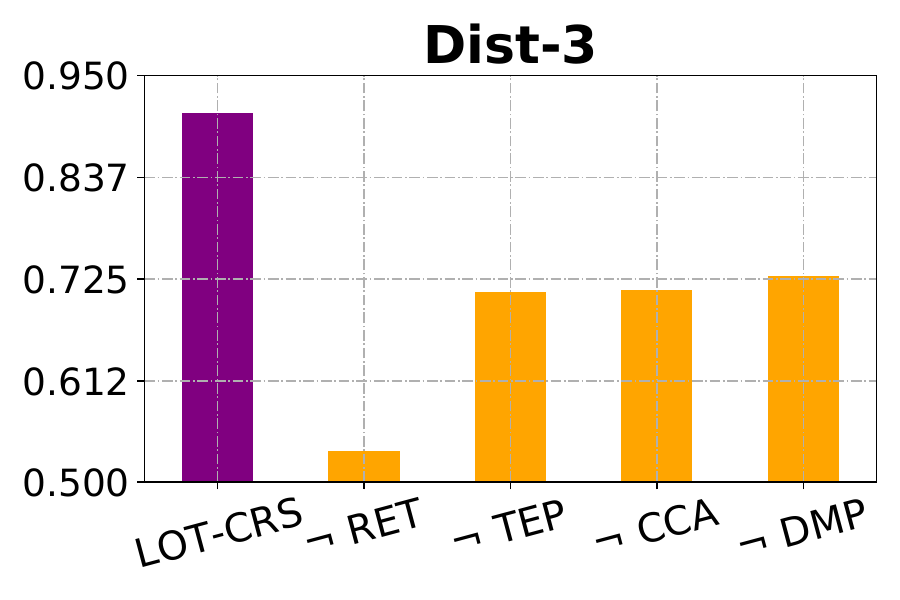}
    \includegraphics[width=0.49\linewidth]{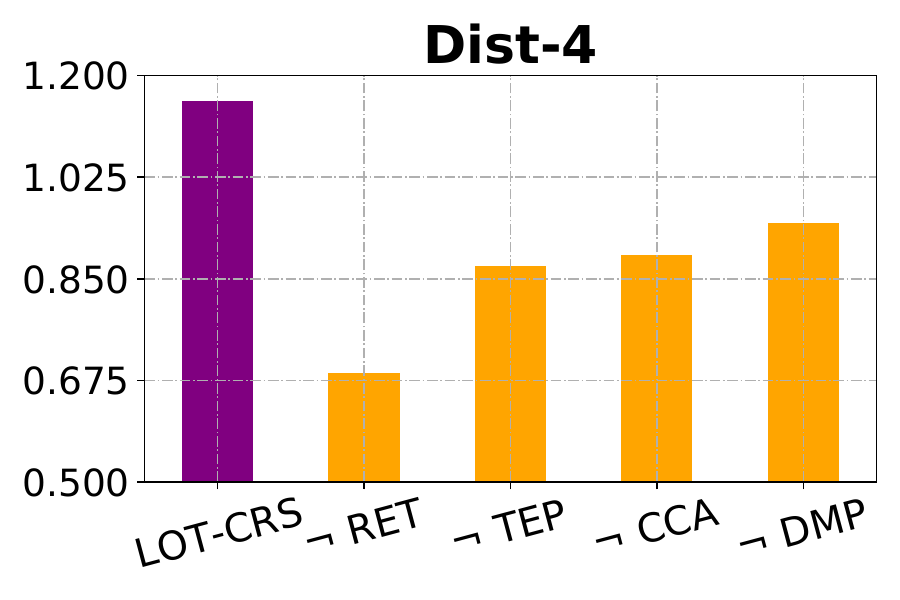}
    \caption{
        Ablation study on the \textsc{ReDial} dataset about the conversation task. 
        RET refers to retrieval-enhanced prompt, TEP denotes response template generation, CCA and DMP denote the two pre-training tasks, \ie contrastive contexts alignment and domain-adaptive masked prediction.}
    \label{fig:ablation-conv}
\end{minipage}

\end{figure*}

\paratitle{Ablation Study.}
We also conduct ablation study on the \textsc{ReDial} dataset, to verify the effectiveness of several components in our approach for the conversation task.
We implement our approach on UniCRS and adopt Distinct-3 and Distinct-4 as the evaluation metrics.
Here, we consider removing the pre-training task DMP and CCA, retrieval-augmented and template generation-and-filling strategies, respectively.
The ablation results are shown in Figure~\ref{fig:ablation-conv}, where RET and TEP refer to the retrieval-augmented and template generation strategies, respectively.
We can see that removing any component would lead to a decrease in the model performance. It shows the effectiveness of all these components in our approach. 
Besides, the retrieval-augmented strategy seems to be more important than others, and it yields a larger performance drop after being removed. This strategy is capable of utilizing the rich informative responses from the simulated CRS datasets, which is helpful for our model to generate more informative responses.

Typically, it requires a considerable amount of training data to learn parameters.
However, in real-world applications, it is likely to suffer from the cold-start issue caused by insufficient data, which may increase the risk of overfitting and make the long-tail problem more serious.
Our approach can effectively alleviate these problems to some extent, with the simulated balanced CRS datasets and specially designed pre-training and retrieval-augmented fine-tuning strategies.
To examine this, we simulate a data scarcity scenario by randomly sampling different proportions of training data.
We select KGSF, DialoGPT and UniCRS for comparison, since they perform  better among the baselines, and implement our approach on UniCRS. 
We report the results of Recall@50 and Tail-Coverage@50 on the \textsc{ReDial} dataset.

Figure~\ref{fig:few-shot} shows the results in different data scarcity settings.
As we can see, the performance of baselines substantially drops by decreasing the available training data, while our method is consistently better than the baselines in all cases.
It indicates that our model can better utilize the limited labeled data and alleviate the cold start problem, especially for the long-tail recommendation.

\subsection{Qualitative Analysis}

To better understand why our approach can lead to performance improvement, we present a sample conversation between a user and our CRS (with UniCRS as the backbone model).

\begin{figure*}[t]
\centering
    \begin{minipage}[c]{0.49\linewidth}
    \centering
    \includegraphics[width=0.6\linewidth]{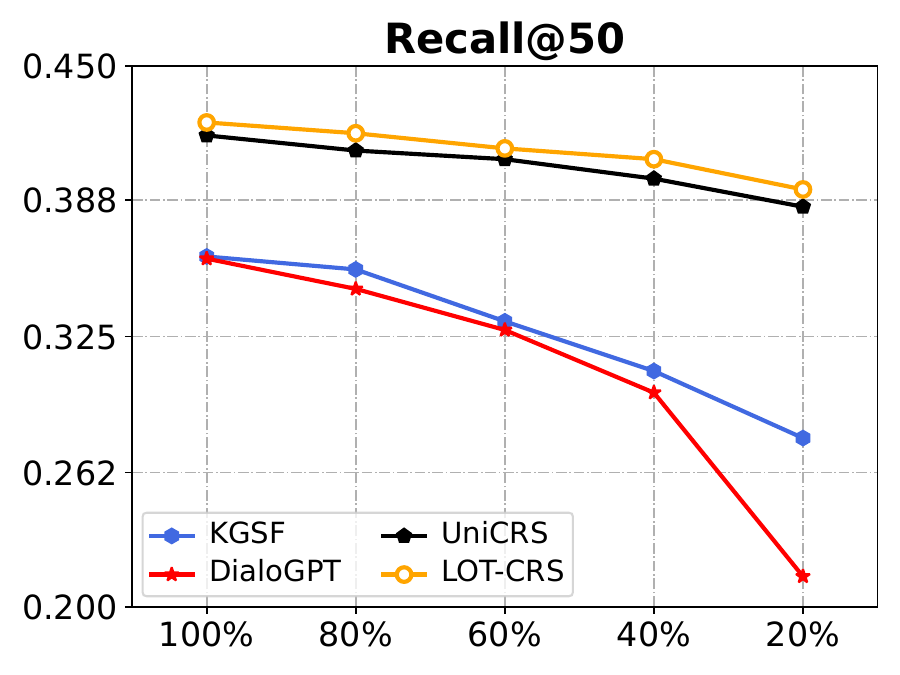}
    \includegraphics[width=0.6\linewidth]{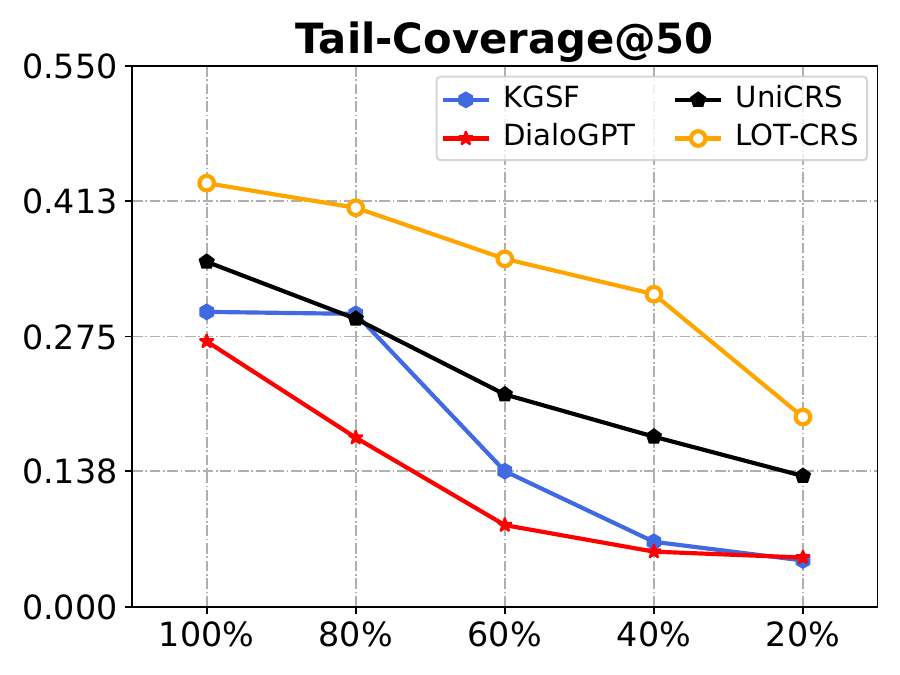}
    \caption{The recommendation performance of our approach \emph{w.r.t.} different amount of training data on \textsc{ReDial} dataset.}
    \label{fig:few-shot}
    \end{minipage}
    \hfill
     \begin{minipage}[c]{0.49\linewidth}
	\includegraphics[width=\linewidth,height=0.9\linewidth]{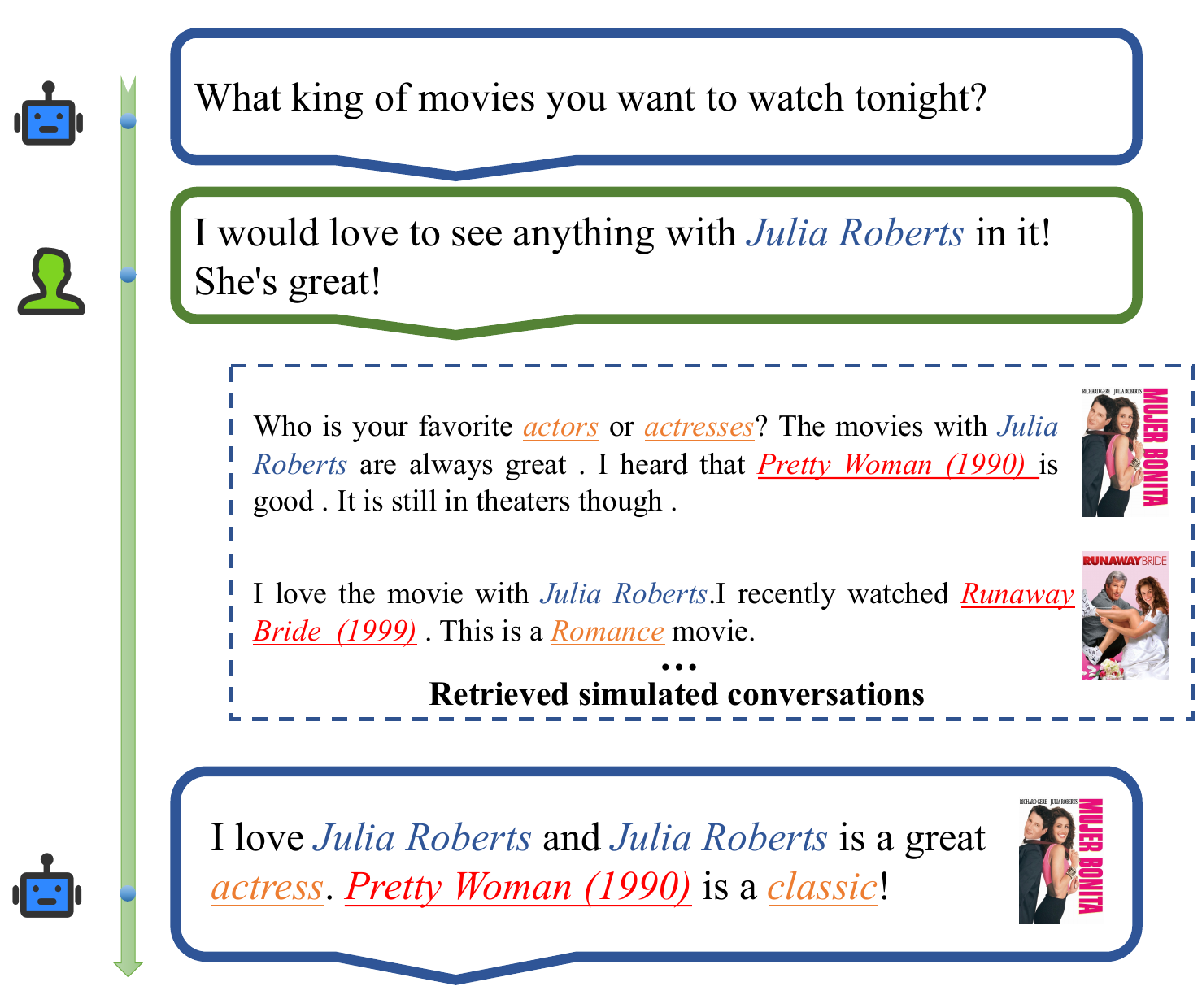}
	\centering
	\caption{A sample conversation between our CRS agent and a real user. We use blue, red and yellow color to mark the entities, items and keywords, respectively.}
	\label{fig:case-study}
     \end{minipage}

\end{figure*}

As shown In Figure~\ref{fig:case-study}, a user is asking our system for recommending   \emph{Julia Roberts}'s movie, which is not the most popular information request in existing CRS datasets (Table~\ref{tab:datasets}). Thus, a CRS trained on such datasets is difficult to make the correct recommendations that meet user's need. 
While, our approach can simulate more balanced conversations covering  the target actress. Further, by performing retrieval on the simulated dataset, our approach can recall a highly relevant simulated converstation, in which a movie \emph{Pretty Woman (1990)} fits the  user's taste well and some important keywords are also included (\eg \emph{actress}).
By referring to the retrieved conversation,  our system successfully recommends a proper movie to the user. 

This case indicates the effectiveness of our approach to alleviating the long-tail problem in conversational recommendation.

